\documentclass[twocolumn,aps,showpacs,prl,superscriptaddress]{revtex4}
\usepackage{graphicx}
\usepackage{dcolumn}
\usepackage{bm}
 \usepackage{amssymb}

\begin{document}

\title{Magnetic quantum oscillations in YBa$_2$Cu$_3$O$_{6.61}$ and YBa$_2$Cu$_3$O$_{6.69}$
in fields of up to 85~T; patching the hole in the roof of the superconducting dome}  

\author{John Singleton}
\affiliation{National High Magnetic Field Laboratory, Los Alamos National Laboratory, MS-E536, Los Alamos, NM 87545, USA}
\author{Clarina de la Cruz}
\affiliation{Department of Physics and Astronomy, The University of Tennessee,
Knoxville, TN~37996-1200, USA}
\affiliation{Neutron Scattering Science Division, Oak Ridge National Laboratory,
Oak Ridge, TN~37831, USA}
\author{R.D. McDonald}
\affiliation{National High Magnetic Field Laboratory, Los Alamos National Laboratory, MS-E536, Los Alamos, NM 87545, USA}
\author{Shiliang Li}
\affiliation{
Institute of Physics, Chinese Academy of Sciences, Beijing 100190, China
}
\affiliation{Department of Physics and Astronomy, The University of Tennessee,
Knoxville, TN~37996-1200, USA}
\author{Moaz Altarawneh} 
\affiliation{National High Magnetic Field Laboratory, Los Alamos National Laboratory, MS-E536, Los Alamos, NM 87545, USA}
\author{Paul Goddard}
\affiliation{Clarendon Laboratory, Department of Physics, Oxford University, Oxford, UK OX1 3PU}
\author{Isabel Franke}
\affiliation{Clarendon Laboratory, Department of Physics, Oxford University, Oxford, UK OX1 3PU}
\author{Dwight Rickel} 
\affiliation{National High Magnetic Field Laboratory, Los Alamos National Laboratory, MS-E536, Los Alamos, NM 87545, USA}
\author{C.H. Mielke} 
\affiliation{National High Magnetic Field Laboratory, Los Alamos National Laboratory, MS-E536, Los Alamos, NM 87545, USA}
\author{Xin Yao}
\affiliation{Department of Physics, Shanghai Jiao Tong University, Shanghai 200030, China}
\author{Pengcheng Dai}
\affiliation{Department of Physics and Astronomy, The University of Tennessee,
Knoxville, TN~37996-1200, USA}
\affiliation{Neutron Scattering Science Division, Oak Ridge National Laboratory,
Oak Ridge, TN~37831, USA}
\affiliation{
Institute of Physics, Chinese Academy of Sciences, Beijing 100190, China
}
\begin{abstract}
We measure magnetic quantum oscillations
in the underdoped cuprates
YBa$_2$Cu$_3$O$_{6+x}$ with $x=0.61$, $0.69$,
using fields of up to 85~T.
The quantum-oscillation frequencies and
effective masses obtained suggest that the Fermi energy in the
cuprates has a maximum at $p\approx 0.11-0.12$.
On either side, the effective mass
may diverge, possibly due to phase
transitions associated with the $T=0$ limit of the metal-insulator
crossover (low-$p$ side), and the postulated topological transition
from small to large Fermi surface close to optimal doping (high $p$ side).
\end{abstract}


\maketitle
One of the significant landmarks in the study of 
the ``High-$T_{\rm c}$" cuprates is 
the observation of Shubnikov-de Haas
and de Haas-van Alphen 
oscillations~\cite{doiron,auduard,bangura,yelland,sebastiannature,bigFS,jaudet}
in high magnetic fields.
Such magnetic quantum oscillations (MQOs)
are the signature of a Fermi surface (FS),
and their temperature ($T$) and field ($B$) dependence suggest a 
relatively conventional Fermi 
liquid~\cite{doiron,auduard,bangura,yelland,sebastiannature,jaudet,sebastian2009},
rendering some theories of the cuprate normal
state untenable~\cite{chakravarty}. 
Though there are attempts to explain the
MQOs using more exotic models~\cite{loughborough,bristol,varma}, 
these seem unable to describe aspects of the
data ({\it e.g.} multiple MQO frequencies,
MQOs periodic in $1/B$, realistic effective masses).

However, published MQOs cover only a restricted region of hole doping $p$.
In particular, data on underdoped YBa$_2$Cu$_3$O$_{6+x}$
correspond to
$0.49\leq x \leq 0.54$ ($0.0925\leq p\leq 0.10$)~\cite{doiron,auduard,sebastiannature,jaudet}.
As this is also the $x$ range blighted by the ortho-I/ortho-II structural
instability~\cite{hardy}, it is natural ask whether the observed FSs
are a consequence of, or related to, this phase separation. Moreover, the
only higher-$p$ data for the underdoped side of the superconducting dome
are for YBa$_2$Cu$_4$O$_8$ ($p\approx 0.125-0.14$~\cite{bangura,yelland}). 
These may be untypical because of the different crystal structure.
Here we therefore report MQOs in the
underdoped cuprates YBa$_2$Cu$_3$O$_{6.61}$ 
($p \approx 0.11$) and YBa$_2$Cu$_3$O$_{6.69}$ ($p \approx 0.125$).
We find that both exhibit a dominant MQO 
frequency $F \approx 550-570$~T, similar to the $\alpha$ frequency
observed in the $0.49\leq x \leq 0.54$
samples~\cite{doiron,auduard,sebastiannature,jaudet,sebastian2009}. 
On close examination, the $p \approx 0.11$ sample exhibits additional
MQO frequencies, some attributable to
warping of the FS due to a finite interlayer transfer integral. 
Effective masses $m^*$ found for both compositions
are less than $2.0m_{\rm e}$,
lighter than their equivalent in YBa$_2$Cu$_4$O$_8$~\cite{bangura,yelland}.
\begin{figure}[h]
\centering
\includegraphics [width=0.90\columnwidth]{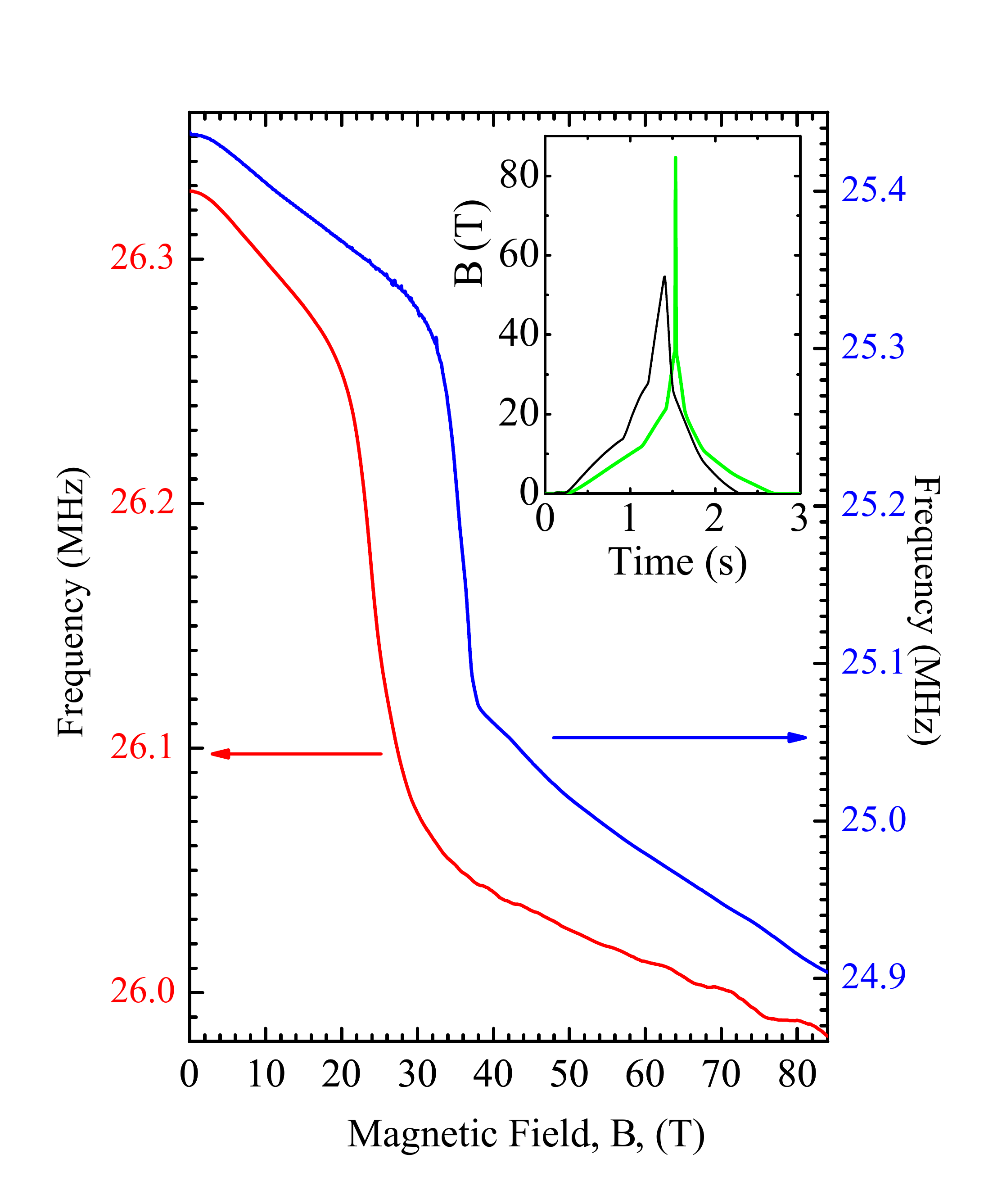}
\vspace{-7mm}
\caption{(color online.) PDC frequency $f$ versus field
$B$ for single crystals of YBa$_2$Cu$_3$O$_{6.69}$
(upper blue trace) and YBa$_2$Cu$_3$O$_{6.61}$
(lower red trace); $T= 1.5$~K for both.
The drop in $f$ corresponds to the
irreversibility field~\cite{yelland,sebastiannature}.
MQOs are clearly
visible at high fields for YBa$_2$Cu$_3$O$_{6.61}$.
The inset shows the $B$ versus time
$t$ profiles for the  60~T Long-pulse and 85~T Multishot magnets.}
\vspace{-8mm}
\label{fig1}
\end{figure}

Single crystals of YBa$_2$Cu$_3$O$_{6.61}$ 
and YBa$_2$Cu$_3$O$_{6.69}$ are grown and oxygenated 
as described before~\cite{growth}. Samples are polished to
sizes $0.3 \times 0.3 \times 1.5$~mm$^3$,
with the long axis parallel to {\bf c}.
Compositions are inferred by
measuring $T_{\rm c}$ in a SQUID magnetometer,
and using the $p$ and $x$ versus $T_{\rm c}$
relationships given in Ref.~\cite{hardy}.
The MQO experiments 
employ the same system as in Ref.~\cite{yelland};
a coil of 8-15 turns of 44 or 50-gauge
Cu wire is wound around the sample,
the planes of the turns roughly
perpendicular to {\bf c}. The
coil is part of a tank circuit driven by either
a tunnel-diode oscillator (TDO)~\cite{mielke}
or a proximity-detector circuit (PDC)~\cite{moaz};
shifts in resonant frequency $f$ are caused by changes in the
skin-depth (normal state) or penetration 
depth (superconducting state)~\cite{mielke}.
No significant differences are noted between 
PDC and TDO data. A heterodyne system measures $f$;
the oscillator output is mixed down
using two mixer/filter stages to about 1~MHz
and the resulting signal digitized directly at $10^7$
samples/s using
a National Instruments PXIe-1062Q
digitizer. Fields are provided by the 85~T Multi-shot
(MSM) and 60~T Long-pulse magnets 
at NHMFL Los Alamos~\cite{yelland,sebastian2009}
and a 65~T short-pulse magnet at Oxford.
The purpose of the range of ${\rm d} B/{\rm d} t$
($\sim 100-15000~{\rm T s^{-1}}$)
is to characterize and eliminate the effects of sample heating
due to induced currents and dissipative vortex motion~\cite{yelland}.
The field is measured using a pick-up coil calibrated
by the belly MQOs of Cu~\cite{goddard}.
Four crystals of YBa$_2$Cu$_3$O$_{6.61}$ 
and two crystals of YBa$_2$Cu$_3$O$_{6.69}$ are studied;
results are consistent between crystals of the same 
$x$ and between different magnets.

Fig.~\ref{fig1} shows data for YBa$_2$Cu$_3$O$_{6.61}$ 
and YBa$_2$Cu$_3$O$_{6.69}$ measured in the 85~T
MSM at $T= 1.5$~K; samples
are heat-sunk to a sapphire chip and immersed in $^4$He liquid.
The frequencies are obtained by Fourier-transforming the
signal using a moving time-window $20~\mu$s long, and then
adding the offset removed by the mixers.
The prominent drop in $f$ around 25~T ($x=0.61$) or 35~T
($x=0.69$) is attributed to the irreversibility field~\cite{yelland,sebastiannature}.
Above this, features are discerned in the
data, corresponding to Shubnikov-de Haas oscillations in
the conductivity~\cite{yelland,sebastiannature}. Owing to the proportionality
between change in conductivity and shift in $f$~\cite{mielke},
the conductivity MQOs give oscillations in $f$.

\begin{figure}
\centering
\includegraphics [width=0.90\columnwidth]{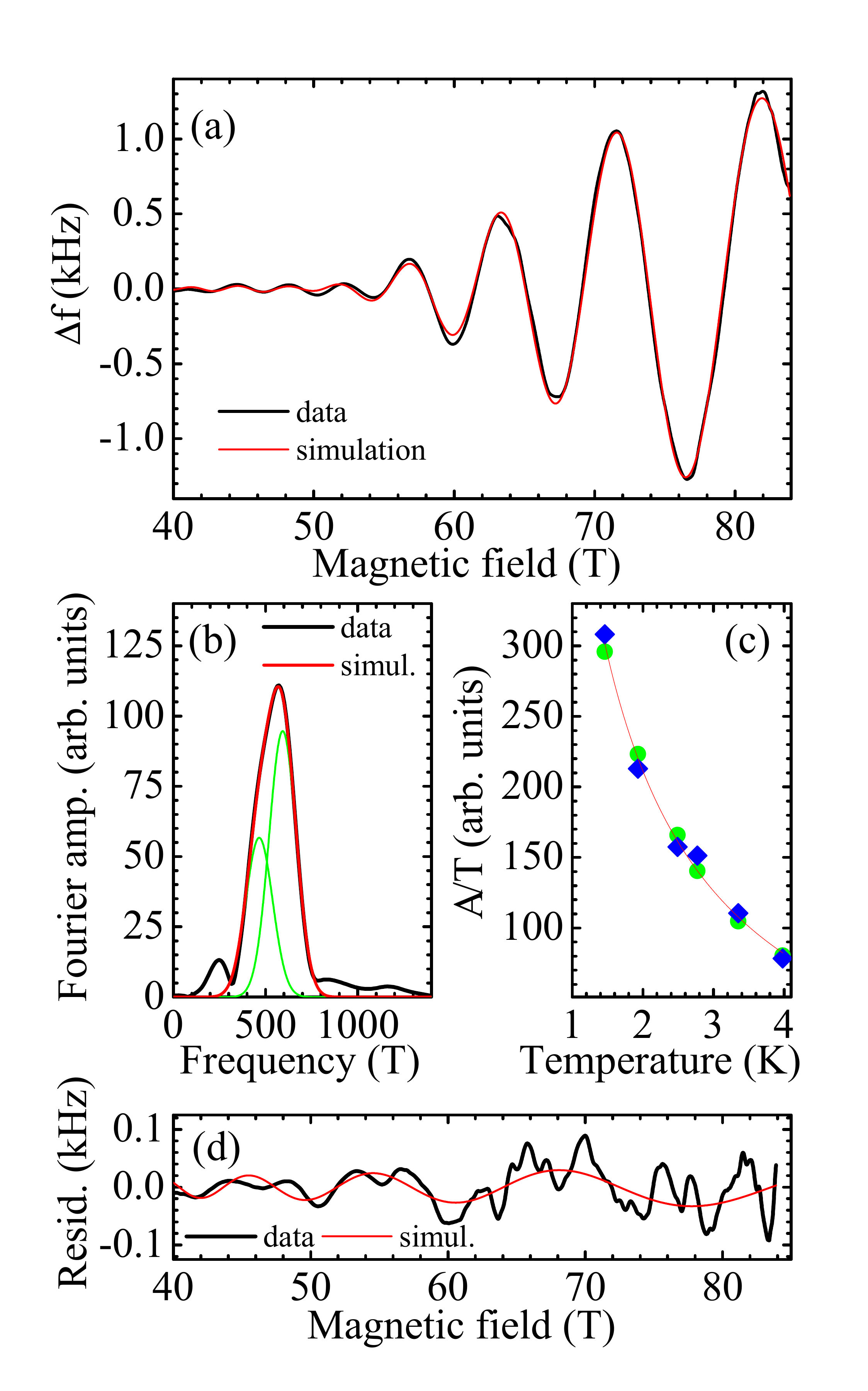}
\vspace{-5mm}
\caption{(color online) (a)~ PDC resonant frequency $f$ for a
YBa$_2$Cu$_3$O$_{6.61}$ crystal after background
subtraction to leave the oscillatory component $\Delta f$;
the trace (thick black curve) is
an average of three magnet sweeps ($T=1.5$~K).
The thinner red line is a fit of Eq.~\ref{ianpaisley} for
MQO frequencies $589$~T and
$479$~T with $\tau \approx 0.07$~ps.
(b)~Fourier transform of the data in (a) (black thick curve);
the large peak is centred on 570~T.
The thinner red curve is the sum of two gaussians
(fine green lines) at 466 and 593~T.
(c)~Plot of MQO amplitude $A$ divided by $T$ versus $T$;
diamonds are from the upsweep of the field
and dots from the downsweep. The curve is a fit of
Eq.~\ref{AoverT}, giving $m^*=1.6\pm 0.1 m_{\rm e}$.
(d)~Residual ({\it i.e.} (data)-(fit)) from (a) versus
field (black thicker line). The thinner red curve
is a fit of Eq.~\ref{ianpaisley} for a single MQO with
$F=270\pm 20$~T.}
\vspace{-5mm}
\label{fig2}
\end{figure}

We first turn to YBa$_2$Cu$_3$O$_{6.61}$
for  which MQOs are visible in the
raw data (Fig.~\ref{fig1}); 
below, we see that MQOs are less
prominent in
YBa$_2$Cu$_3$O$_{6.69}$ due to
a shorter apparent scattering time, $\tau$.
To make the MQOs more visible, the slowly-varying
background due to the semiclassical magnetoresistance is removed
by subtracting a third-order polynomial in $B$.
Figure~\ref{fig2}(a)  shows some resulting MQOs
for YBa$_2$Cu$_3$O$_{6.61}$;
here, random noise from the power supply of the
85~T MSM~\cite{sebastian2009} is mitigated by averaging
three upsweeps and three downsweeps and then smoothing
using a Savitsky-Golay routine. The resulting
data exhibit MQOs above about $40$~T.
On Fourier-transformation, the dominant peak is at
$F=570$~T (Fig.~\ref{fig2}(b)),  
similar to the so-called $\alpha$ MQO frequency ($500-550$~T) in 
$0.49\leq x\leq0.54$ samples~\cite{doiron,auduard,sebastiannature,sebastian2009}
and the dominant frequency in YBa$_2$Cu$_4$O$_8$~\cite{bangura,yelland}.

However, in the case of a single extremal FS cross-section,
one would expect the MQO amplitude to grow uniformly with increasing
field~\cite{shoenberg}. The MQOs in Fig.~\ref{fig2}(a) do not do this;
they are modulated by what appears to be a beat frequency, a
phenomenon noted other cuprates~\cite{yelland,auduard,ross}.
This is also seen in the Fourier transform (Fig.~\ref{fig2}(b)),
where the peak at 570~T 
is obviously asymmetric and may be fitted by two overlapping Gaussians
centred on $466\pm 10$ and $593\pm 5$~T.
The presence of two relatively closely-spaced
Shubnikov-de Haas frequencies with similar amplitudes is
suggestive of the beats caused by ``neck and belly''
oscillations of a quasi-two-dimensional FS that
is warped due to a finite interlayer transfer integral $t_{\bf c}^{\perp}$~\cite{auduard,ross}.
To model this, we sum two components of the 
Lifshitz-Kosevich formula~\cite{ross,shoenberg} with
MQO frequencies $F_1$ and $F_2$, amplitudes $a_1$ and $a_2$
and phases $\phi_1$ and $\phi_2$:
\[
\Delta f = \left(a_1\cos\left[\frac{2\pi F_1}{B}+\phi_1\right]+
a_2\cos\left[\frac{2\pi F_2}{B}+\phi_2\right]\right)\times
\]
\vspace{-7mm}
\begin{equation}
TB^{-\frac{1}{2}}\exp\left[-\frac{\pi m^*}{e\tau B}\right]\left(\sinh \left[\frac{14.69 m^*T}{B}\right]\right)^{-1}
\label{ianpaisley}
\end{equation}
Here $m^*$ is the effective mass,
and $\tau^{-1}$ is an effective scattering rate;
we assume that $m^*$ and $\tau$ are the same for the neck and belly oscillations.
Independently, $m^*$ may be constrained by the way
in which the amplitude of an individual MQO, or the Fourier amplitude
of a transform over a restricted field range varies with $T$;
\begin{equation}
\frac{A}{T} \propto \left(\sinh \left[\frac{14.69 m^*T}{B}\right]\right)^{-1}
\label{AoverT}
\end{equation}
where $A$ is the amplitude.
All of the fits ({\it e.g.} Fig.~\ref{fig2}(c))
of individual MQOs or Fourier amplitudes yielded
$m^*$ values in the range $1.5-1.7m_{\rm e}$,
irrespective of sample, field range or sweep rate,
leading us to $m^*=1.6 \pm 0.1 m_{\rm e}$.
Having constrained $m^*$,
a fit of Eq.~\ref{ianpaisley} to the data (Fig.~\ref{fig2}(a))
yields MQO frequencies $589 \pm 5$~T and
$479\pm 5$~T, and $\tau \approx 0.07$~ps.
These values are close to those obtained
in the two-gaussian fit of the transform in Fig.~\ref{fig2}(b).

Beside the peak at 570~T in the transform
(Fig.~\ref{fig2}(b)),
there is a feature at $F\approx 250$~T. This appears to correspond to 
an actual MQO series, as is seen by
subtracting the fitted Eq.~\ref{ianpaisley} in Fig.~\ref{fig2}(a)
from the data. The residual (Fig.~\ref{fig2}(d)) is oscillatory,
with a direct fit yielding $F=270\pm 20$~T, close 
to the value suggested by the peak
in the Fourier transform~\cite{susan}. Unfortunately, the MQOs are
too poorly defined to permit estimates of 
$m^*$ or $\tau$.

To summarize for YBa$_2$Cu$_3$O$_{6.61}$,
our data suggest three FS
cross-sections, with MQO frequencies
270~T, 479~T and 589~T; other peaks in the 
transform at higher frequencies (Fig.~\ref{fig2}(b)) are attributable
to harmonics of these~\cite{disclaimer}. The
479~T and 589~T MQOs are likely
the neck and belly oscillations of a warped quasi-two-dimensional
FS, with $m^*=1.6\pm 0.1 m_{\rm e}$;
this is probably the equivalent of the dominant $\alpha$ frequency
in other underdoped cuprates~\cite{doiron,auduard,sebastiannature,sebastian2009}.
The frequency difference, $\Delta F \approx 110$~T, between neck
and belly oscillations
suggests~\cite{ross} an average interlayer transfer integral
$t_{\bf c}^{\perp}=\hbar \Delta F/(4m^*) = 2.0 \pm 0.1$~meV for YBa$_2$Cu$_3$O$_{6.61}$,
higher than the values $1.4-1.7$~\cite{spread} obtained for 
YBa$_2$Cu$_3$O$_{6+x}$ ($x=0.51$, 0.54)~\cite{auduard,ross}.
This increase in $t_{\bf c}^{\perp}$ with
$p$ is not unexpected; the lattice parameter
{\bf c} declines with $p$~\cite{hardy}.

\begin{figure}
\centering
\includegraphics [width=0.98\columnwidth]{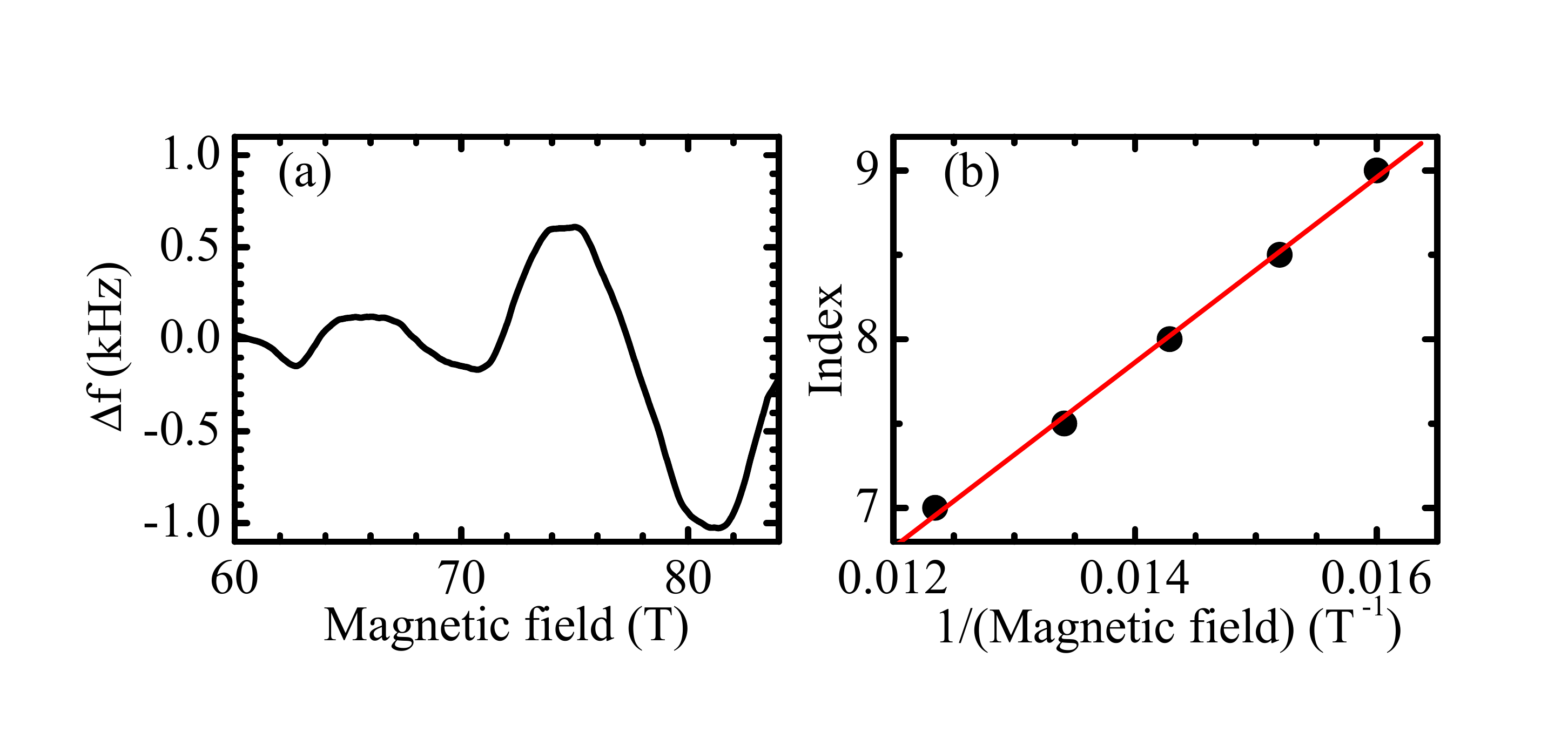}
\vspace{-8mm}
\caption{(color online) (a)~ PDC resonant frequency for a
YBa$_2$Cu$_3$O$_{6.69}$ crystal after background
subtraction to leave the oscillatory component $\Delta f$;
the trace is a smoothed  average of three magnet sweeps ($T=1.5$~K).
(b)~Oscillation index versus reciprocal magnetic
field for the MQOs in (a) (points); dips in $\Delta f$ are indexed
by integers and peaks by half integers. The straight line is
a fit with a gradient of $550 \pm 20$~T.}
\vspace{-6mm}
\label{fig3}
\end{figure}

\begin{figure}
\centering
\includegraphics [width=0.95\columnwidth]{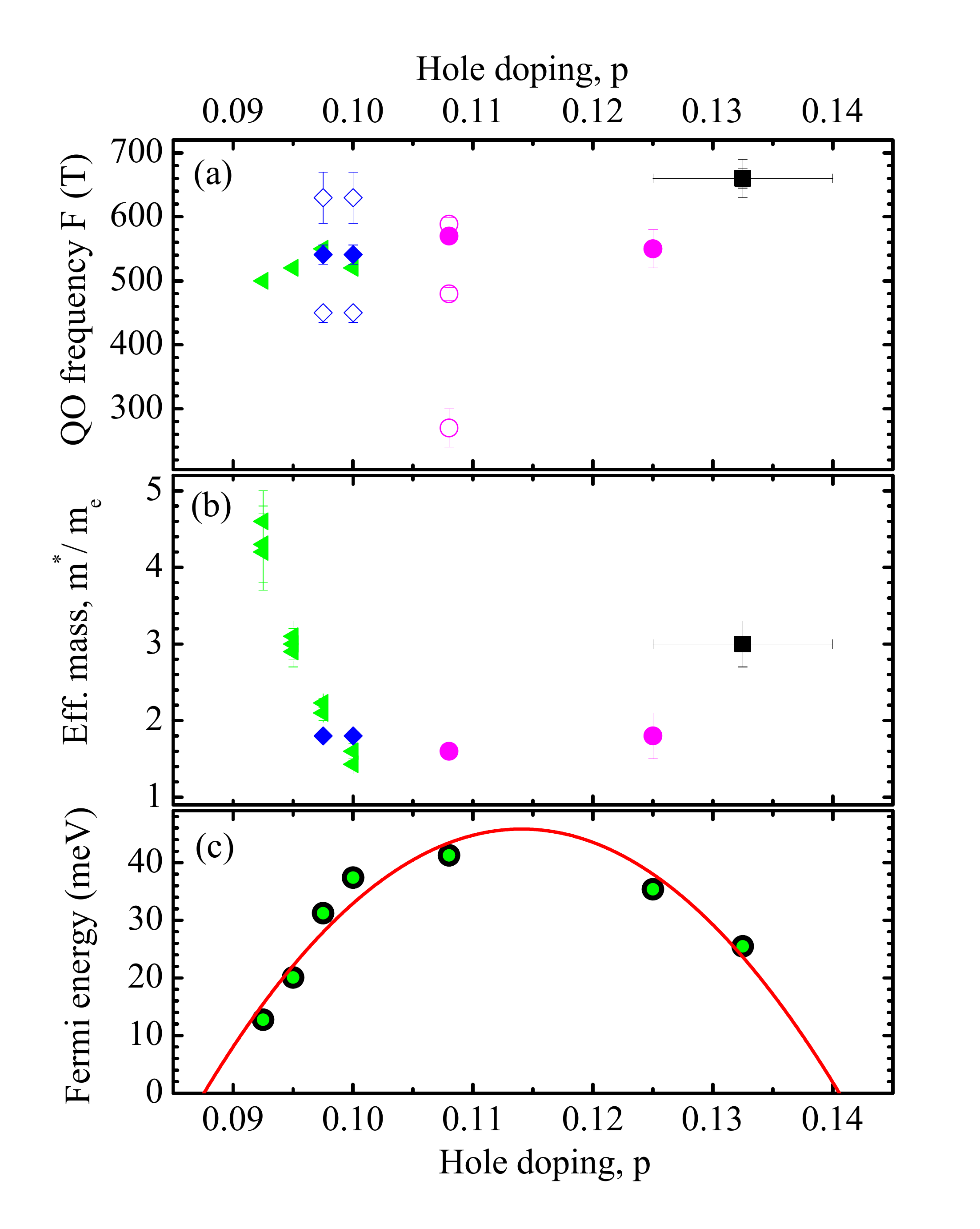}
\vspace{-8mm}
\caption{(color online) (a)~ Summary of MQO
frequencies versus $p$ for underdoped cuprates:
for YBa$_2$Cu$_3$O$_{6+x}$,
$\lhd$ are from Ref.~\cite{sebastian2009},
$\Diamond$ from Ref.~\cite{auduard}, and $\circ$ from
this work; $T_{\rm c}$ and $x$ values are
converted to $p$ using Ref.~\cite{hardy}. 
YBa$_2$Cu$_4$O$_8$ data from Refs.~\cite{bangura,yelland}
are squares; the horizontal bar is the
spread in $p$ values given for 
YBa$_2$Cu$_4$O$_8$~\cite{bangura,yelland}. 
Solid symbols ({\it e.g.} $\bullet$) show the dominant ($F_\alpha$) 
frequency obtained from Fourier
analysis; open symbols are from more detailed analysis
({\it e.g.} Figs.~\ref{fig2}(a,d) or Refs.~\cite{ross,auduard}). 
(b)~Effective mass of the dominant ($\alpha$) MQO
frequency $F_{\alpha}$ versus $p$; symbols are the 
same as in (a) except $\Diamond$ are from Ref.~\cite{jaudet}.
(c)~Fermi energy from $F_{\alpha}$ and $m^*$; for
$p$s where several values are given, we take the average.
Points are data and the curve is a parabolic fit.}
\vspace{-6mm}
\label{fig4}
\end{figure}

Figure~\ref{fig3}(a) shows an example of the MQOs
observed in YBa$_2$Cu$_3$O$_{6.69}$.
In contrast to YBa$_2$Cu$_3$O$_{6.61} $,
where MQOs appear around 40~T (Fig.~\ref{fig2}(a)),
the MQOs here are not distinguishable from the background
until about 60~T (Fig.~\ref{fig3}(a)). The non-sinusoidal appearance
of the MQOs again suggests the presence of more than one frequency,
but sadly, the limited field window over which MQOs 
are seen both precludes a ``neck and belly'' analysis 
(Eq.~\ref{ianpaisley}, Fig.~\ref{fig2}(a)) and limits the
resolution of a Fourier transform. Instead, we plot MQO index
versus $1/B$ in Fig.~\ref{fig3}(b) 
to find a mean frequency of $550\pm 20$~T~\cite{footnote}.
Fitting the MQO amplitudes versus $T$
for YBa$_2$Cu$_3$O$_{6.69}$ yields $m^* = 1.8 \pm 0.3 m_{\rm e}$,
similar to the $1.6\pm 0.1 m_{\rm e}$ 
for the analogous MQO frequency
in the YBa$_2$Cu$_3$O$_{6.61}$ (Fig.~\ref{fig2}(c)).
A Dingle analysis ({\it i.e.} a plot of $\log_e(AB^{\frac{1}{2}}\sinh(14.69m^*T/B)$ versus
$1/B$~\cite{shoenberg}, where $A$ is the oscillation 
amplitude) yields $\tau \approx 0.04$~ps, 
$\sim 2$ times smaller than that 
for YBa$_2$Cu$_3$O$_{6.61}$.
This accounts for the higher fields required to observe
MQOs in YBa$_2$Cu$_3$O$_{6.69}$.
Ref.~\cite{us} attributes the dominant Landau-level broadening
to quasistatic spin disorder
also observed in neutron experiments
and parameterized by a correlation length $\xi$~\cite{mcd,kampf,yamada}.
It is notable that $\xi$ decreases with increasing $p$~\cite{mcd,kampf,yamada},
and this may account for the shorter $\tau$ of the $x=0.69$ samples.

Figure~\ref{fig4} compares the data obtained here with
similar results from other underdoped cuprates, all of
which have a dominant MQO frequency
$F_{\alpha} \approx 500-660$~T. 
Figure~\ref{fig4}(a) shows both $F_{\alpha}$
and  other MQO frequencies $\leq~1000$~T
that have been resolved (this paper, Refs.~\cite{auduard,ross}).
If we attribute $F=540$ and $450$~T
for $p\approx 0.10$~\cite{auduard} and $F=590$ and
$480$~T for $p=0.11$ (this work) to the
belly and neck oscillations of the $\alpha$ Fermi pocket,
then there seems to be a trend,
continued by YBa$_2$Cu$_3$O$_8$,  for 
the $\alpha$ pocket to grow with rising $p$.
It also seems that samples from the orthoI-II region
are unexceptional, continuing the
trend seen in this work to lower $p$.
The weaker MQOs with $F=630$~T ($p = 0.975$, 0.10)~\cite{auduard}
$F=270$~T ($p=0.11$) are qualitatively similar
to extra pockets predicted by FS 
reconstruction due to various types of symmetry breaking;
{\it e.g.} an
incommensurate spin-density wave~\cite{neilFS}
produces a plethora of FS sheets, both smaller and larger than
the $\alpha$ pocket, whilst a pocket with $F\approx 250$~T is
an explicit prediction of incommensurate d-density-wave
order~\cite{dimov,chakravartykee}.
Meanwhile, the $\alpha$ effective masses show a 
``bowl''-shaped dependence on $p$, with
a minimum at $p\approx 0.11$.

To vizualize effect that these changes
have on the carrier system, Fig.~\ref{fig4}(c) plots
the effective Fermi energy $E_{\rm F}$ for the $\alpha$ pocket,
$E_{\rm F} = \hbar F_{\alpha}/m^*$,
using data from Figs.~\ref{fig4}(a,b).
It seems that the 
Fermi energy reaches a maximum
at $p \approx 0.115$, but decreases either side of this,
suggesting that $m^*$ may diverge at $p \approx 0.087$
and $p\approx 0.14$, the latter $p$ being
poorly constrained by the existing data~\cite{point}.
The lower $p$ value suggests the point at which
Metal-Insulator transition
tends to $T=0$~\cite{sebastian2009,nr11}. 
The upper may signal the topological transition
from  small to large FS thought to occur close to optimal
doping~\cite{yelland,bigFS}, though experimental
confirmation of an unreconstructed FS in
overdoped YBa$_2$Cu$_3$O$_{6+x}$ is as yet lacking.
By analogy with heavy-fermion
superconductors~\cite{sebastian2009,hfs}, both of the
$m^*$ divergences may represent quantum
critical phase transitions.

In summary, we report MQO frequencies
and effective masses for the underdoped cuprates
YBa$_2$Cu$_3$O$_{6+x}$ with $x=0.61$, $0.69$,
filling in a considerable gap in the FS versus
hole doping diagram.
In conjunction with other data, our results
suggest that the Fermi energy
may reach a maximum around $p\approx 0.11-0.12$,
and collapse on either side due to divergence of the effective mass.
The divergences are perhaps associated with quantum-critical phase
transitions associated with the $T=0$ limit of the metal-insulator
transition (low-$p$ side), and the topological transition
from small to large FS close to optimal doping (high $p$ side).

This work is supported by the DoE BES 
grant ``Science in 100~T'', DOE DE-FG02-05ER46202, and 
in part by Division of Scientific User Facilities.  
NHMFL is funded by DoE, NSF and the State of Florida.
Work at the IOP is supported by the Chinese Academy of Sciences.
SJTU is supported by Shanghai Committee of Science 
and Technology and the MOST of China (2006CB601003). 
Work at Oxford takes place
in the Nicholas Kurti Magnetic Field Laboratory and
is supported by EPSRC.
We thank Neil Harrison for useful discussions and
John Betts, Mike Gordon, Alan Paris, Daryl Roybal and Chuck
Swenson for extreme technical assistance.


\begin{thebibliography}{99}
\bibitem{spread}
The spread in possible values is due to the slightly different fit procedures
and $\Delta F$ values
in Refs.~\cite{auduard,ross} and the range of effective masses
given in the literature.
\bibitem{point}
The conflicting $p$ values~\cite{bangura,yelland} for YBa$_2$C$_4$O$_8$
are the the chief cause of the uncertainty.
\end{thebibliography}
\end{document}